# On calculations of dipole moments of HCl$^+$ and DCl$^+$ molecular ions

V.S. Gurin[1], M.V. Korolkov[2]

[1] Research Institute for Physical Chemical Problems, Belarusian State University, Minsk, Belarus

[2] A.V. Stepanov Institute of Physics, National Academy of Sciences, Minsk, Belarus

**Abstract**

Dipole moment functions of isotopomeric molecular ions, HCl$^+$ and DCl$^+$, are considered in the two coordinate systems, center of mass of nuclei and center of nuclear charges, both through simple analytical derivations and *ab initio* calculations of electronic structure at various interatomic separations. An origin of the different values for dipole moments of the isotopomers is discussed and demonstrated by the calculation data.

## 1. Introduction

Electronic structure calculations within the framework of conventional Born-Oppenheimer approximation provide identical results for molecules of isotopomers, as far as the electronic shell of isotopes are the same. E.g., the electronic structure of HCl and DCl have no difference, and major quantum-chemical methods and corresponding software do assume the Born-Oppenheimer approximation that is quite suitable for many chemical tasks [1]. However, some physical properties of molecules depend on nuclear mass, and therefore isotopomers appear to be different. Vibrational properties of molecules are one such example, that is commonly known and trivial. More interesting question regards dipole moment (DM) function of charged molecules since the position of center of mass of molecules, evidently, depends on mass of isotopes, and that fact changes value of molecular property, in particular, DM. Major quantum-chemical calculations use the system of center of nuclear charges in processing of data, in which case the factor of nuclear mass is lost. In the present work, using the data on *ab initio* calculation of electronic structure of HCl$^+$ (DCl$^+$) molecular ions we consider how DM of HCl$^+$ and DCl$^+$ can be different depending on type of coordinate system used for analysis.

## 2. DM definitions

The general definition of DM includes both electronic and nuclear parts:

$$D_{full} = D_N + D_e \qquad (1)$$

There are two commonly used coordinate systems to treat molecular properties: the center of mass of nuclei (CNM) and the center of nuclear charges (CNC). The first one is mostly used in various quantum mechanical tasks as well in any other areas of physics in which mass of constituent particles is important. The latter originated from the Born-Oppenheimer principle widely used in quantum-chemical treatments through separation of electronic and nuclear parts in the Schroedinger equation. This separation results in the principal contribution of the electronic part in major chemical tasks, and nucleus masses do not enter quantum chemical calculations explicitly. The masses became to be important in solutions associated with nuclear motion, e.g. vibrational spectrum, molecular dynamic, etc. If we treat, for example HCl molecule in the term of quantum-chemical task, we use as parameters the nuclear charges, $Z_N(H) = 1$ $e$ and $Z_N(Cl) = 17$ $e$, where $e$ is the elementary charge. No any atomic masses are used in the calculations, that is quite understandable and does not rise the problems until we work with one type of nuclei for a chemical element (H here). Within the framework of this 'chemically standard' approach, HCl and DCl possess the *same* electronic structure, and as consequence, the same properties determined by the electronic structure and DM of HCl and DCl became the same. Meanwhile, even purely classical treatment of the diatomic system with different masses of atoms, evidently, leads to different values of DM [LL,   ]. The classical value should be true at large separations of atoms when electronic structure factors have minimum contributions. An apparent discrepancy of intuitive and more rigorous calculations of DM of the diatomics issues from the difference in the coordinate systems used.

For molecular ions, in contrast to neutral molecules, the situation acquires a particular significance since DM value of ions depends on the coordinate system. In the case of neutral particles there is no matter what is a reference point for evaluation of DM, and the latter reflects the genuine property of a molecule. However, in the case of a non-zero full charge the reference point position varies DM value under consideration. This can be shown easily if to use the definition of DM with no respect to the type of constituents of the system [2]:

$DM = \Sigma e_i \mathbf{r}_i,$

where the sum is taken through all i charges and **r** are corresponding radius-vectors. $e_i$ are charges of constituents. In different coordinate systems shifted on arbitrary constant vector **a**

$$\mathbf{r}_i' = \mathbf{r}_i + \mathbf{a} \tag{2}$$

then

$$DM' = \Sigma e_i \mathbf{r}_i' = \Sigma e_i \mathbf{r}_i + \mathbf{a}\Sigma e_i = DM + \mathbf{a}\Sigma e_i. \tag{3}$$

If the full charge of the system is zero, i.e. $\Sigma e_i = 0$, DM' = DM, but otherwise, if $\Sigma e_i \neq 0$, the formula (3) simply results that DM' ≠ DM since **a** is considered also to be not zero for any meaningful transformations (2).

Thus, two coordinate systems described above for molecular calculations, CNM and CNC, if appear not the same (e.g. when a nuclear mass of two molecular ions is different), result in DM' ≠ DM. Later, we derive relationships for DM in the CNC and CNM systems and demonstrate using the results of *ab initio* calculations the numerical difference between values of DM for $HCl^+$ and $DCl^+$.

### 3. Analytical derivations

Let us consider how the full DM, relation (1), behaves in CNC and CNM. In CNC the first term, evidently, $D_N = 0$, but in CNM this is not so, and the difference for DM of HCl and DCl appears. We use common formulas for CNM and CNC assuming the 1-dimensional case with all distances only along the molecular axis. Then the differences in values of DM are due to the differences in position of origins of CNM and CNC. Below we derive relations for them.

$$r_1 + r_2 = R, \tag{4}$$

where $r_1$ and $r_2$ are distances to atoms in CNM system (Fig. 1a), and R is the interatomic distance. Usual definitions read through masses of the atoms $m_1$ and $m_2$:

$$r_1 = (m_2/(m_1+m_2))R \text{ and } r_2 = (m_1/(m_1+m_2))R, \tag{5}$$

Analogously, for CNC we use distances $r_{z1}$ and $r_{z2}$ to atoms with respect to this CNC system (Fig. 1b) and R is the same, atom-atom separation:

$$r_{z1} + r_{z2} = R, \qquad (6)$$

$$r_{z1} = (z_2/(z_1+z_2))R \text{ and } r_{z2} = (z_1/(z_1+z_2))R, \qquad (7)$$

$z_1$ and $z_2$ are nuclear charges those define the center of nuclear charges in the same way as masses $m_1$ and $m_2$ give the center of mass.

Now, from these formulas let us evaluate the difference between position of CNC and CNM with (5) and (7): $r_{z2}-r_2$ or $r_{z1}-r_1$ that give the same final result:

$$r_{z2} - r_2 = (z_1/(z_1+z_2))R - (m_1/(m_1+m_2))R \qquad (8)$$

Then for HCl $z_1=17$, $z_2=1$, $m_1=35$, $m_2=1$ and we have $(17/18 - 35/36)R = -(1/36)R =$
**-0.027(7)R.** **(9)**

For DCl $z_1=17$, $z_2=1$, $m_1=35$, $m_2=2$ results in $(17/18 - 35/37)R =$
**-0.0015(015)R.** **(10)**

These additions for values of DM should be used to pass from calculated values in CNC, i.e. typical quantum-chemical calculations, to the results in CNM (molecular mechanics and dynamics).

Now, the differences in one coordinate system, CNM, can be easily derived for DCl and HCl:

$$r_{2(DCl)} - r_{2(HCl)} = \left[\frac{m_{1(HCl)}}{m_{1(HCl)} + m_{2(HCl)}} - \frac{m_{1(DCl)}}{m_{1(DCl)} + m_{2(DCl)}}\right] *R =$$

$$\frac{m_{1(HCl)}m_{2(DCl)} - m_{1(DCl)}m_{2(HCl)}}{[(m_{1(HCl)} + m_{2(HCl)})][(m_{1(DCl)} + m_{2(DCl)})]} *R = -35/(36*37)R = \qquad (11)$$

**= -0.026(276)R**

If to evaluate this relation with $r_1$ instead of $r_2$, i.e. $r_{1(DCl)} - r_{1(HCl)}$ the result will be same.

## 4. Ab initio calculations

The calculations for the ground and several lower states of $HCl^+$ and $DCl^+$ were presented in our recent works [3-5]. For the present analysis of DM values we use here the SCF-HF level calculation with basis set of 6-311G(2DF,2PD). We concern the $A^2\Sigma^+$ state, and for other ones the application is analogous. This state is featured by the asymptotic of DM(R) with monotonous increase and can easily demonstrate the idea of the present study.

Fig. 2 displays the results of *ab initio* calculations of DM for this state of $HCl^+$ and $DCl^+$ ($A^2\Sigma^+$) in the different coordinate systems, CNC and CNM, and the numerical data are collected in Table 1 together with the value

$$DD = (1/R)(DM(CNM) - DM(CNC)) \qquad (12)$$

demonstrating how the formula (10) for difference in the coordinate systems under consideration is valid in the calculations. The data in CNC are coincident for both ions, in CNM the difference is maximum for short distances.

DM function plotted in Fig. 2 clearly show the difference between the value for $HCl^+$ and $DCl^+$ cases, and it grows at the asymptotic, in accordance with analytical derivations above. The asymptotic behavior is described well by the classical values evaluated from the masses of elements, 35/37 and 36/37.

## 5. Conclusion

Using analytical derivations and quantum-chemical *ab initio* calculation for DM functions of isotopomeric molecular ions, $HCl^+$ and $DCl^+$, we consider the origin of difference of DM values that is noticeably increased with R to classical asymptotics. This occurs in the CNM coordinate system rather than in CNC (used in most calculations).

Table 1. Numerical data of calculations for HCl$^+$ and DCl$^+$

| R, $a_0$ | DM, DCl$^+$, CNM | DM, HCl$^+$, CNM | DM, DCl$^+$, CNC | DM, HCl$^+$, CNC | DD, DCl |
|---|---|---|---|---|---|
| 1.5 | 0.017448 | 0.05686241 | 0.01519574 | 0.01519574 | 0.00150151 |
| 2 | 0.38422686 | 0.43677941 | 0.38122386 | 0.38122386 | 0.0015015 |
| 2.5 | 0.83032879 | 0.89601948 | 0.82657503 | 0.82657503 | 0.0015015 |
| 3 | 1.22289816 | 1.30172699 | 1.21839366 | 1.21839366 | 0.0015015 |
| 3.5 | 1.56539419 | 1.65736116 | 1.56013893 | 1.56013893 | 0.0015015 |
| 4 | 1.9125054 | 2.0176105 | 1.90649939 | 1.90649939 | 0.0015015 |
| 4.5 | 2.34191625 | 2.46015949 | 2.33515949 | 2.33515949 | 0.0015015 |
| 5 | 3.01164677 | 3.14302815 | 3.00413926 | 3.00413926 | 0.0015015 |
| 5.5 | 4.0475866 | 4.19210612 | 4.03932835 | 4.03932835 | 0.0015015 |
| 6 | 5.02022976 | 5.17788742 | 5.01122075 | 5.01122075 | 0.0015015 |
| 6.5 | 5.75779053 | 5.92858632 | 5.74803077 | 5.74803077 | 0.0015015 |
| 7 | 6.36348451 | 6.54741845 | 6.352974 | 6.352974 | 0.0015015 |
| 7.5 | 6.90619138 | 7.10326346 | 6.89493012 | 6.89493012 | 0.0015015 |
| 8 | 7.41866598 | 7.62887619 | 7.40665397 | 7.40665397 | 0.0015015 |
| 8.5 | 7.91614435 | 8.1394927 | 7.90338159 | 7.90338159 | 0.0015015 |
| 9 | 8.4058542 | 8.64234069 | 8.39234069 | 8.39234069 | 0.0015015 |
| 9.5 | 8.89127804 | 9.14090267 | 8.87701378 | 8.87701378 | 0.0015015 |
| 10 | 9.37415169 | 9.63691445 | 9.35913668 | 9.35913668 | 0.0015015 |
| 10.5 | 9.8553217 | 10.1312226 | 9.83955594 | 9.83955594 | 0.0015015 |
| 11 | 10.33525389 | 10.62429293 | 10.31873737 | 10.31873737 | 0.0015015 |
| 11.5 | 10.8142412 | 11.11641837 | 10.79697393 | 10.79697393 | 0.0015015 |
| 12 | 11.29246413 | 11.60777944 | 11.27444611 | 11.27444611 | 0.0015015 |
| 12.5 | 11.77005736 | 12.09851082 | 11.75128859 | 11.75128859 | 0.0015015 |
| 13 | 12.24712272 | 12.58871431 | 12.2276032 | 12.2276032 | 0.0015015 |
| 13.5 | 12.72373998 | 13.07846971 | 12.70346971 | 12.70346971 | 0.0015015 |
| 14 | 13.1999733 | 13.56784117 | 13.17895228 | 13.17895228 | 0.0015015 |
| 14.5 | 13.67587525 | 14.05688125 | 13.65410347 | 13.65410347 | 0.0015015 |
| 15 | 14.15148942 | 14.54563356 | 14.1289669 | 14.1289669 | 0.0015015 |
| 15.5 | 14.62684336 | 15.03412564 | 14.60357008 | 14.60357008 | 0.0015015 |
| 16 | 15.1019866 | 15.52240702 | 15.07796257 | 15.07796257 | 0.0015015 |
| 16.5 | 15.57693598 | 16.01049454 | 15.5521612 | 15.5521612 | 0.0015015 |
| 17 | 16.05171375 | 16.49841045 | 16.02618823 | 16.02618823 | 0.0015015 |
| 17.5 | 16.52631069 | 16.98614553 | 16.50003441 | 16.50003441 | 0.0015015 |
| 18 | 17.00080111 | 17.47377408 | 16.97377408 | 16.97377408 | 0.0015015 |
| 18.5 | 17.47517171 | 17.96128282 | 17.44739393 | 17.44739393 | 0.0015015 |
| 19 | 17.94943367 | 18.44868292 | 17.92090514 | 17.92090514 | 0.0015015 |
| 19.5 | 18.42359784 | 18.93598522 | 18.39431856 | 18.39431856 | 0.0015015 |
| 20 | 18.89767384 | 19.42319936 | 18.86764381 | 18.86764381 | 0.0015015 |
| 20.5 | 19.37167015 | 19.91033382 | 19.34088937 | 19.34088937 | 0.0015015 |
| 21 | 19.84559428 | 20.39739608 | 19.81406275 | 19.81406275 | 0.0015015 |
| 21.5 | 20.31945285 | 20.88439279 | 20.28717057 | 20.28717057 | 0.0015015 |
| 22 | 20.79325176 | 21.37132984 | 20.76021873 | 20.76021873 | 0.0015015 |
| 22.5 | 21.26699624 | 21.85821245 | 21.23321245 | 21.23321245 | 0.0015015 |
| 23 | 21.74069097 | 22.34504533 | 21.70615644 | 21.70615644 | 0.0015015 |
| 23.5 | 22.21434014 | 22.83183264 | 22.17905486 | 22.17905486 | 0.0015015 |
| 24 | 22.68794751 | 23.31857814 | 22.65191148 | 22.65191148 | 0.0015015 |
| 24.5 | 23.16151645 | 23.80528522 | 23.12472966 | 23.12472966 | 0.0015015 |
| 25 | 23.63505 | 24.29195691 | 23.59751247 | 23.59751247 | 0.0015015 |
| 25.5 | 24.10855092 | 24.77859596 | 24.07026263 | 24.07026263 | 0.0015015 |
| 26 | 24.58202167 | 25.26520486 | 24.54298263 | 24.54298263 | 0.0015015 |

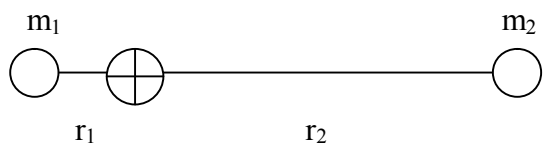

Fig. 1a A scheme for definition of values of r and m in CNM

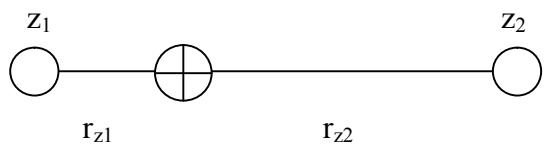

Fig. 1b A scheme for definition of values of r and z in CNC

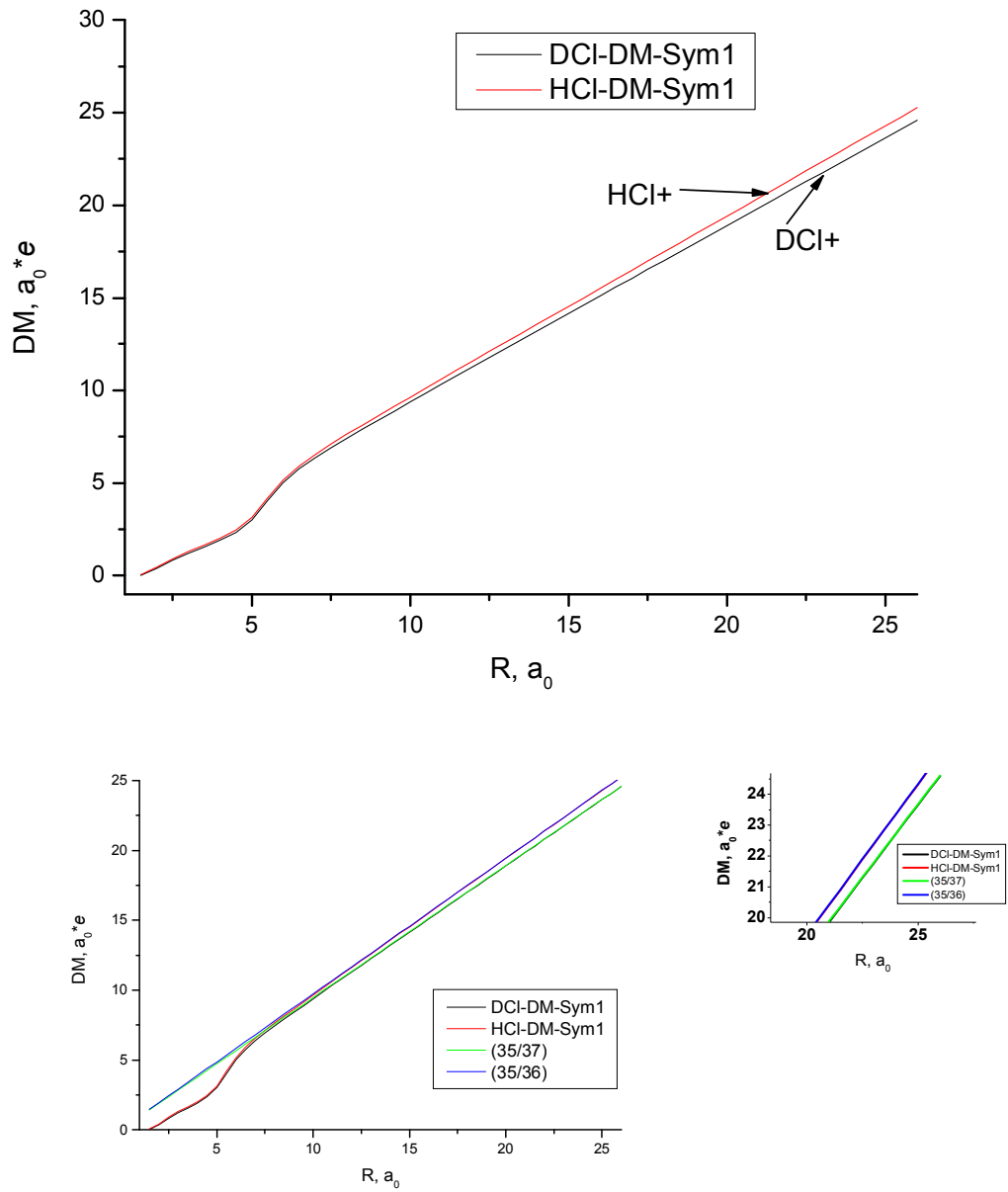

Fig. 2. Plots of the calculation results for $HCl^+$ and $DCl^+$ in CNM and the same with comparison of lines $y=(35/37)x$ and $y=(35/36)x$ to view the correspondence with analytical results above.

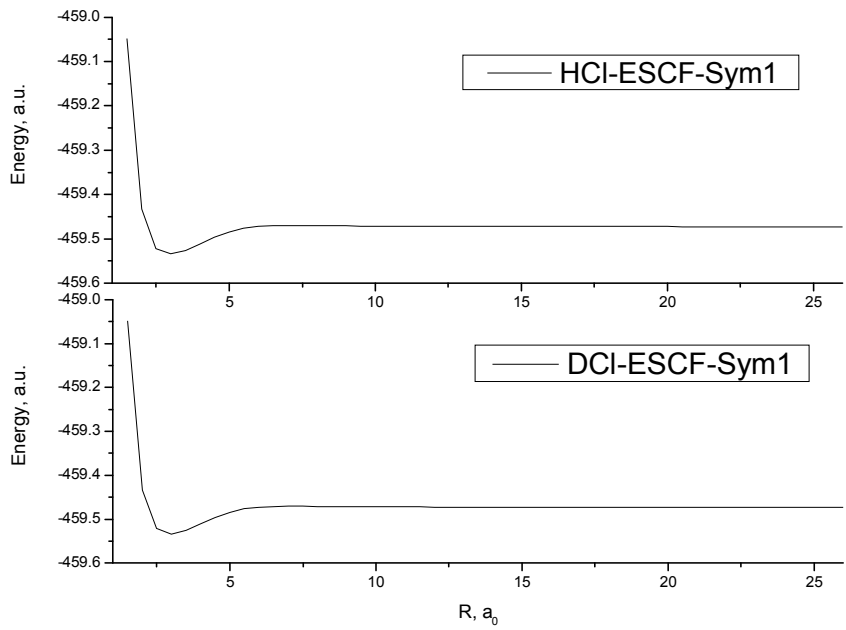

Fig. 3. Potential curves of $A^2\Sigma^+$ of $HCl^+$ and $DCl^+$.